\documentclass{article}

\usepackage{arxiv}

\usepackage[utf8]{inputenc} 
\usepackage[T1]{fontenc}    
\usepackage{hyperref}       
\usepackage{url}            
\usepackage{booktabs}       
\usepackage{amsfonts}       
\usepackage{nicefrac}       
\usepackage{microtype}      
\usepackage{cleveref}       
\usepackage{lipsum}         
\usepackage{graphicx}
\usepackage[square,numbers]{natbib}
\usepackage{doi}

\usepackage[caption=false]{subfig}

\usepackage{todonotes}

\begin{document}

\title{How Large Language Models Are Changing MOOC Essay Answers: A Comparison of Pre- and Post-LLM Responses}

\author{
 Leo Leppänen\\
  University of Helsinki\\
  Helsinki, Finland \\
  \texttt{leo.leppanen@helsinki.fi} \\
  \And
 Lili Aunimo\\
  Haaga-Helia University of Applied Sciences\\
  Helsinki, Finland\\
  \texttt{lili.aunimo@haaga-helia.fi} \\
  \And
 Arto Hellas\\
  Aalto University\\
  Espoo, Finland \\
  \texttt{arto.hellas@aalto.fi} \\
  \And
 Jukka K. Nurminen\\
  University of Helsinki\\
  Helsinki, Finland \\
  \texttt{jukka.k.nurminen@helsinki.fi} \\
  \And
 Linda Mannila\\
  University of Helsinki\\
  Helsinki, Finland\\
  \texttt{linda.mannila@helsinki.fi}\\
}
\maketitle              
\begin{abstract}
The release of ChatGPT in late 2022 caused a flurry of activity and concern in the academic and educational communities. Some see the tool's ability to generate human-like text that passes at least cursory inspections for factual accuracy ``often enough'' a golden age of information retrieval and computer-assisted learning. Some, on the other hand, worry the tool may lead to unprecedented levels of academic dishonesty and cheating. In this work, we quantify some of the effects of the emergence of Large Language Models (LLMs) on online education by analyzing a multi-year dataset of student essay responses from a free university-level MOOC on AI ethics. Our dataset includes essays submitted both before and after ChatGPT's release. We find that the launch of ChatGPT coincided with significant changes in both the length and style of student essays, mirroring observations in other contexts such as academic publishing. We also observe---as expected based on related public discourse---changes in prevalence of key content words related to AI and LLMs, but not necessarily the general themes or topics discussed in the student essays as identified through (dynamic) topic modeling. 

\end{abstract}
%
%
%

\section{Introduction}


Languages and vocabularies have always evolved through, for instance, cultural development, historical events, and social interactions~\cite{campbell2021}. Today, however, large language models (LLMs) and tools like ChatGPT are influencing languages in new ways~\cite{reviriego2024}. In addition to affecting current language use, this may impact the future of languages, as new models are trained on AI-generated content \cite{martinez2023} and individuals will learn languages while increasingly exposed to AI-generated text \cite{reviriego2024}. 
AI-based changes in vocabulary can already be observed, e.g., as LLMs tend to favor certain words over others \cite{kobaketal2024}. 

The quick development within generative AI (GAI) has led to fast uptake of and easy access to LLM-based tools. For instance, ChatGPT had the fastest user growth in Internet history until July 2023, reaching one million users in just five days \cite{duarte2024}. According to OpenAI CEO Sam Altman, ChatGPT had 100 million weekly users less than a year after its launch. The number of monthly visits at ChatGPT.com varies from 152 million in November 2022 to its current number of visits (as of December 2024), which is 4.97 billion, ranking eighth among the world's most visited websites.  The tools are used as writing aids by different user groups, quite naturally also among university students \cite{tossell2024student}. 

Recent research has highlighted the influence of the emergence of LLMs on academic texts~\cite{geng2024,kobaketal2024}, job applications~\cite{kauttonen2024topic}, and student-produced content~\cite{kauttonen2024topic,safi2023work,denny2023can}. Work analyzing differences between LLM- and human-produced content in course contexts has primarily focused on specifically created datasets, with observations of changes in coursework over time having received little attention. 

In this article, we analyze student essay responses to a free MOOC (massive open online course) on ethics of AI that has been in operation since late 2020. To our knowledge, this is the first study investigating language changes in MOOC essays surrounding the introduction of LLMs using longitudinal data. We focus on quantifying the impact of the emergence of LLM-based tools, studying differences in essay responses before and after the release of ChatGPT. Our overarching question for the present work is: \textit{How have essays submitted by students changed following the release of ChatGPT and the broader emergence of LLM-based tools?}

\section{Background}


\subsection{Impact of Technology on Everyday Language}\label{tech-impact-language}

Societal changes affect language and vocabulary, which are shaped both by how we interact with the world~\cite{ferreiro2019} and through historical and societal events, such as wars and pandemics \cite{bochkarev2014universals,kobaketal2024}. Waves of technological progress have also had a similar impact; as an example, the industrial revolution gave rise to technical terms describing new machinery and innovations, many of which became lasting additions to our dictionaries. 

While early communication technologies, such as print and broadcasting, also introduced new conventions, the influence of the Internet is different due to speed and reach. Dictionaries regularly update definitions and add terms originating from online culture. As an example, the word ``selfie'' was first used in 2002, with Oxford Dictionaries naming it the international word of the year in 2013 \cite{word-of-the-year}. In addition to creating new words, technology also repurposes existing ones, such as \textit{friend}, \textit{like}, \textit{wall} and \textit{block}, which have taken on new meanings in digital contexts \cite{reed2014}. Similar evolutions can be seen in acronyms, where, for instance, \textit{LOL, ``Laughing Out Loud''} first signaled amusement, but is now interpreted more as sarcasm. Digital platforms have also affected how we express ourselves, both stylistically and in terms of brevity. For instance, character limits, such as those set by SMS and Twitter, led to more condensed writing and increased use of emojis and acronyms, condensing multi-word expressions into single terms integrated into everyday communication~\cite{tagliamonte2016}. 

Linguistic adaptation to technological advances is ongoing. In recent years, AI-related jargon, such as machine-learning, neural networks, bias and LLMs, has become part of public discourse. Previous shifts in AI terminology, such as, for instance, from traditional machine learning (such as decision trees, support vector machines, and linear regression) to deep artificial neural networks, have mainly mainly concerned experts. GAI development, however, extends beyond technical communities and affects society as a whole, with shifting perceptions of what qualifies as AI, typical use cases and related technologies~\cite{bo2025}. Terminology used in AI often evolves from technical jargon to widely adopted metaphors, creating challenges when terms lose their original context, gain unintended meanings, or reinforce misleading narratives about the capabilities and impact of AI~\cite{rehak2021}. Another shift in AI-related terminology can be seen as a widening from primarily purely technical concepts to encompassing societal concerns related to, e.g., ethics, privacy, copyright, trustworthiness and sustainability~\cite{almenzar2022}. 

As AI becomes embedded in everyday life, perceptions of what qualifies as AI, its use cases, and related technologies will shift. The AI effect~\cite{haenlein2019} suggests that as technology successfully solves a problem and its solution becomes widely adopted, it is no longer considered AI. Such conceptual changes are also reflected in vocabulary, but understanding of the meaning behind words and expressions may be limited, as is commonly the case when technical subjects enter public discourse~\cite{sentance2021}. The same vocabulary may hence be understood and used differently by experts and non-experts \cite{garcia2023}. 

\subsection{Stylistic Changes in Language Due to GAI}

Recent studies have documented measurable changes in word frequency distributions, lexical diversity, and syntactic characteristics, highlighting how GAI and LLMs can influence written language.

Kobak et al. \cite{kobaketal2024} analyzed PubMed abstracts after the launch of ChatGPT. Looking into yearly word frequencies in abstracts, their findings highlight an increase in the use of certain words, such as \emph{delve, crucial, potential, these, significant}, while the word \emph{important} has become less common. The language produced by LLMs also evolve. As an example, looking into GPT-3.5 and GPT-4, Reviriego et al.~\cite{reviriego2024} found a shift from a more limited vocabulary with lower lexical diversity compared to humans, to a similar -- or surpassing -- level of lexical diversity to human language.



Word frequencies have also been used to study the use of LLMs in scientific writing. For instance, Geng and Trotta \cite{geng2024} studied word frequency changes in 1 million arXiv abstracts from 2018 to 2024 and found that LLMs appear to be increasingly used, in particular in computer science. Liang et al. \cite{liang2024} came to the same conclusion in a study of 950k papers. Similarly, Diamond~\cite{diamon2023} studied whether word frequency distributions in AI-generated texts follow Zipf's law, which states that word frequency is inversely proportional to rank in a corpus. 
The results showed that AI-generated languages also closely adhere to Zipf's law, suggesting that they share statistical properties with human languages.

In addition to vocabulary and word frequencies, several other features have been found to correspond with AI authorship in scientific writing. André et al.~\cite{andre2023detecting} compared linguistic and stylometric characteristics of human-written and AI-generated texts. The features studied were perplexity, grammar, n-gram distributions, function word distributions, Type-Token Ratios (TTR) and token length. In their study, they found that AI-generated texts tend to exhibit lower perplexity, less grammatical errors, a less diverse ngram distribution, a lower TTR ratio, longer tokens and a narrower distribution of function words.

Furthermore, readability metrics like Flesch Reading Ease score and the associated Flesch-Kincaid grade level, which assess text complexity, have also been used to study stylistic differences between AI and human-generated texts~\cite{mindner2023classification}. Some metrics have also been developed for texts to be learnt by ESL (English as a foreign language) students \cite{uchida-etal-2024-profiling}. 

More broadly, research has also compared other linguistic aspects including topic shifts, textual density, and contextual cues. For instance, studies have shown that in code-related tasks, AI tends to generate shorter code but longer explanations when compared to humans~\cite{denny2023can}. TF-IDF and average word length measurements have also been used to identify distinct patterns between student-produced texts and those generated by AI, where students have tended to use shorter responses than AI~\cite{safi2023work}; this may, however, be language specific~\cite{yildiz2025comparison} and also relate to the English proficiency of the text author~\cite{mizumoto2024identifying}. There is also differing levels of involvement and quality, where AI-generated texts have ``artificial quality'', while human-generated texts tend to have higher degree of narrativity and more context-specific references~\cite{sardinha2024ai}, and more personal references~\cite{goulart2024ai}. 

While studies analyzing scientific literature have used broader corpora such as manuscripts from arXiv, studies of differences between student- and AI-generated essays often focus on a sample of essays collected specifically for the purposes of the study. In this work, we analyze the potential broader influence of GAI on student essays, highlighting changes in student-submitted coursework.

\section{Methodology}

\subsection{Context}
Ethics of AI is a free open online course developed by the University of Helsinki and its partners, aimed at anyone interested in the ethical dimensions of AI. The 2 ECTS credit\footnote{European Credit Transfer System -- 1 ECTS credit= approximately 27 hours of study.} course focuses on building understanding of what AI ethics entails, exploring the boundaries of ethical AI development, and fostering an ethical mindset in AI-related decision-making. 
The course covers topics such as non-maleficence, accountability, transparency, human rights, fairness, and practical applications of AI ethics, and provides a comprehensive overview of the field. 
The course materials are available at \url{https://ethics-of-ai.mooc.fi/}.

\subsection{Research Questions and Data}

To investigate the changes in student essays, we address the following research questions: \textit{How have student-submitted essays changed following the release of ChatGPT, in terms of RQ1) style and length of writing, RQ2) non-topic related vocabulary, and RQ3) essay topics?}

 


The study uses essay submissions from 3,582 course participants, taking the course in English, who consented to their data being used for research. The essays were peer-reviewed by other students, who had the option to mark submissions as spam or otherwise unsuitable. In addition, course staff monitored the submissions. During data pre-processing, we omitted all essays that had been flagged as spam or as otherwise unsuitable by either students or course staff. After filtering, our data consisted of a total of 56,878 English-language essay submissions. The earliest submission included in the dataset was from November 2020, and the latest from October 2024.

Due to data minimization and privacy, student demographics are not available for the present study. Due to the global nature of the course, we expect them to include participants from all over the globe and from varying backgrounds, with a significant proportion (if not majority) being non-native English speakers.

\subsection{Analysis}

We split the essays into three time periods: data prior to the release of ChatGPT in November 2022 (21,869 essays); data in the first year after the release (i.e., November 2022--November 2023; 12,782 essays); and data from at least one year after the release (i.e., December 2023 or later; 22,227 essays). In the following analyses, we focus on the two timespan extremes, as we expected the middle timespan to show a slow increase in ChatGPT usage. As the later analyses will show, this method of dividing the data into segments was more conservative than what strictly would have been necessary.

In order to answer RQ1, we first examine the lengths of the essay submissions based on tokens and sentences, sentence lengths, and the difficulty of the essays as measured by Flesch Reading Ease. We use NLTK's~\cite{bird2006nltk} \texttt{sent\_tokenize} and \texttt{word\_tokenize} functions to process the essays into tokens and sentences. 

To answer RQ2, we investigate the changes in prevalence of some of the key terms identified by Kobaketal et al.~\cite{kobaketal2024}. Following André et al.~\cite{andre2023detecting}, we also calculate the Type-Token Ratio of the essays. When calculating the prevalence of terms, we include words with both capital and non-capital first letters, and for words other than ``these'', we allow arbitrary suffixes. For example, the term ``crucial'' also includes the token ``Crucially''. We report only relative prevalences, i.e., for each essay we calculate the number of occurrences of the term and divide by the length of the essay in tokens, and then average these numbers across all essays. To determine whether a statistically significant difference exists in relative prevalences between two time periods, we use the Mann-Whitney U test.

To answer RQ3, we use two methods. First, we developed a list of words and terms associated with GAI tools and LLMs, such as ``Generative AI'', ``ChatGPT'', ``large language model'' etc.; a list of ``classical'' AI terminology such as ``knowledge representation'', ``NLP'', ``neural networks'', etc.;  and a list of terms relating to ethical aspects of AI, such as ``bias'', ``deep fake'', ``power consumption'' etc. We then investigated how the relative prevalence of these terms changed between the pre-ChatGPT essays and those submitted at least one year post-ChatGPT. We ignore most irrelevant capitalization and some of the more common suffixes such as a word-final `s' in `language model', and allow for some common spelling and capitalization variations. For example, the term ``GPT-3.5'' also includes ``gpt3'', but ``RAG'' does not include ``rag''. 
Furthermore, we apply Gensim topic models (both ``normal'' and dynamic) to essays relating to two specific essay prompts to determine whether there are any larger-scale changes in the topics over the time spans.

\section{Results}

\subsection{Style and Length of Writing}

\begin{figure}[]
\centering
\subfloat[Token count\label{fig:token_count}]{%
  \includegraphics[width=0.49\textwidth]{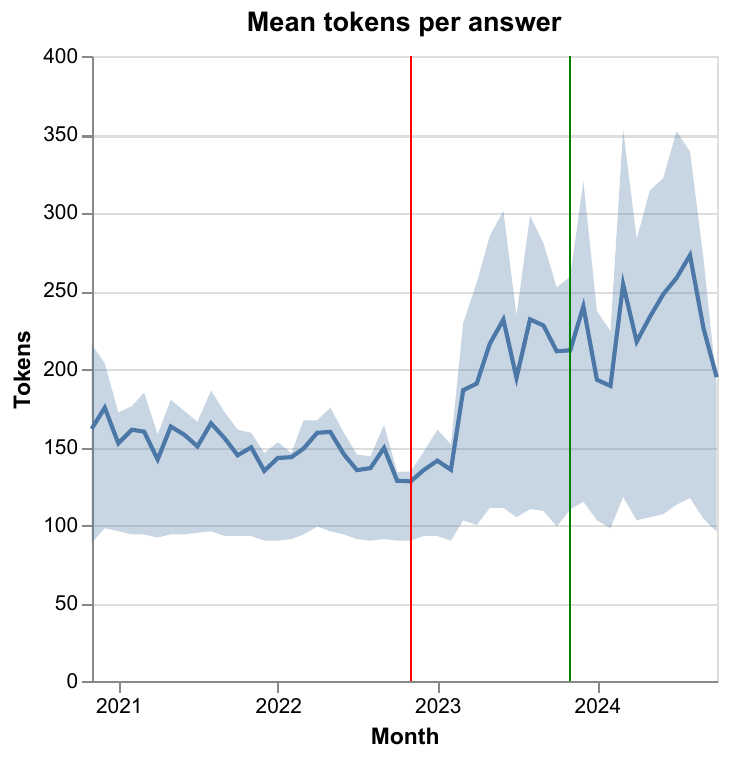}%
}\hfil
\subfloat[Sentence count\label{fig:sent_count}]{%
  \includegraphics[width=0.49\textwidth]{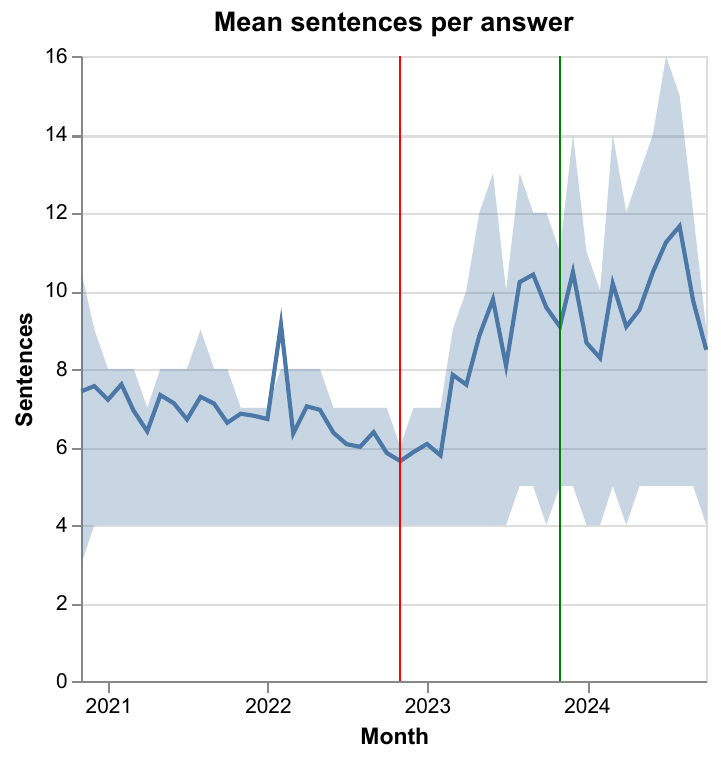}%
}\\
\subfloat[Flesch Reading Ease\label{fig:flesch}]{%
  \includegraphics[width=0.49\textwidth]{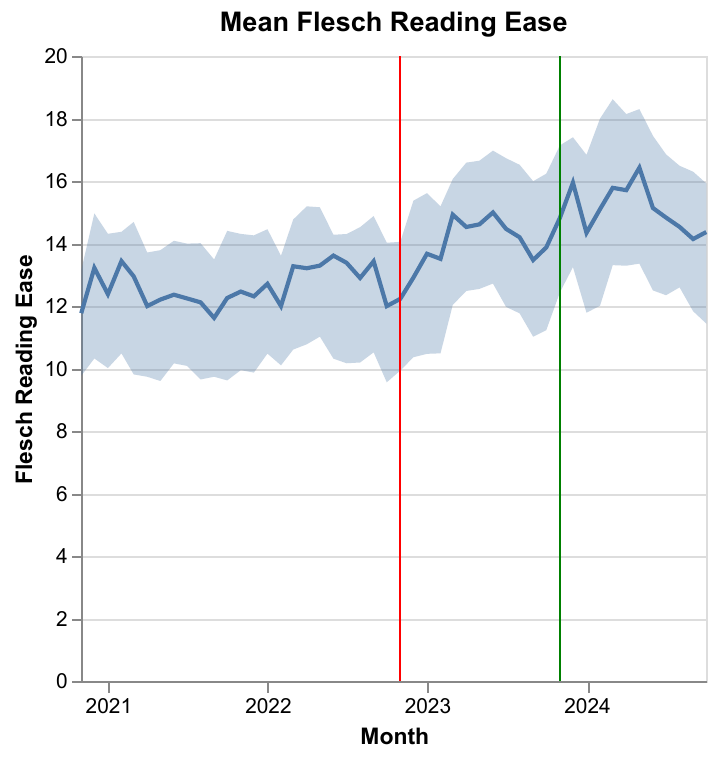}%
}\hfil
\subfloat[Type-Token Ratio\label{fig:ttr}]{%
  \includegraphics[width=0.49\linewidth]{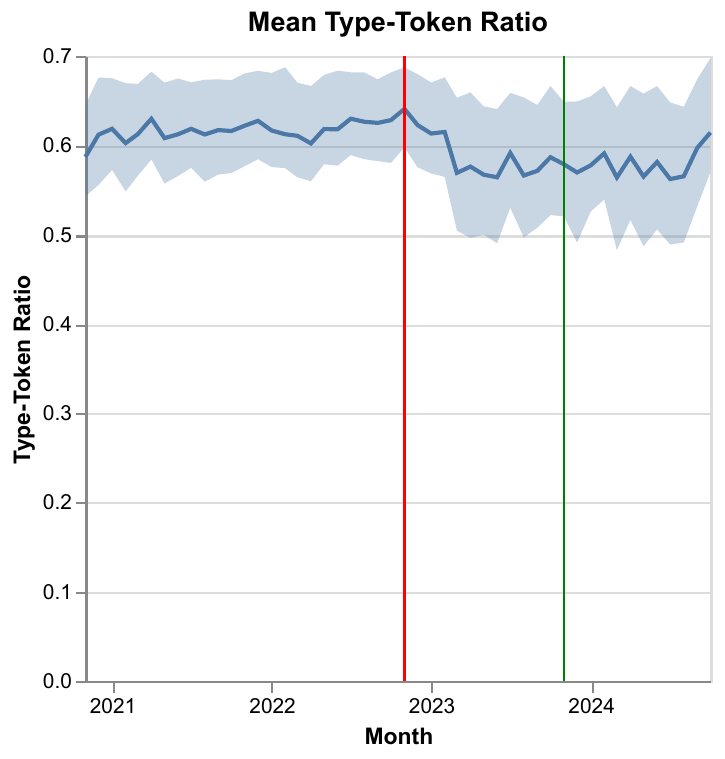}
}
\caption{Key text statistics. Line indicates mean, and the shaded area indicates the first and third quartiles of the data. Data in monthly bins. The first vertical line indicates the release of ChatGPT, and the second vertical line indicates one year after the release.}
\label{fig:lengths}
\end{figure}

Figure~\ref{fig:token_count} shows the evolution of mean answer length measured in tokens in monthly bins. The data shows an aggressive jump around March 2023. The difference between pre-ChatGPT answers (mean 150.5 tokens, std. 104.3) and those submitted at least one year post-ChatGPT (mean 230.1 tokens, std. 171.0) is statistically significant (MWU, U=165056438.0) at p < 0.0001. 

Figure~\ref{fig:sent_count} shows the evolution of sentence counts. We observe an equivalent time lag, before the sentence count surges around March 2023. The difference between pre-ChatGPT answers (mean 6.85, std. 5.82) and answers submitted at least one year post-ChatGPT (mean 9.76, std. 7.76) is statistically significant (MWU, U=177093486.5) at p < 0.0001.

At the same time, while we do observe a statistically significant change in sentence lengths (MWU U=223322456.5, p < 0.0001), the change is minute in magnitude with means rising from 24.91 (std. 11.56) tokens per sentence pre-ChatGPT to 25.56 (std. 17.91) at least one year post-ChatGPT. 

These changes correspond with a statistically significant (MWU U=53227275.5, p < 0.0001) change in the Flesch Reading Ease (Figure~\ref{fig:flesch}), which increases from a mean of 12.70 (std. 4.84) pre-ChatGPT to a mean of 15.31 (std. 7.72) at least one year post-ChatGPT. Recall that a higher Flesch Reading Ease score indicates text that is \textit{easier} to read, meaning we observe a decrease in text difficulty. College graduate texts are expected to fall between 30 and 10. The point of change is not quite as easy to identify from this graph, but the score first increases above 14 in early 2023, mirroring our previous observations.

\subsection{Non-Topic Related Variance}

In terms of the type-token ratio of the essays (Figure~\ref{fig:ttr}), our results mirror those of André et al.~\cite{andre2023detecting}, in that we observe a statistically significant (MWU U=296354583.5, p < 0.0001) drop in the type-token ratio from a mean of 0.617 (std. 0.091) pre-ChatGPT to a mean of 0.577 (std. 0.113) at least one year post-ChatGPT. This indicates that the texts are becoming less varied in terms of vocabulary used within the individual texts.


We also investigated several words that previous works indicate are commonly used by LLMs. The evolution in the relative prevalence of these words is shown in Table~\ref{tab:llm-words}. We observe statistically significant increases in the use of LLM-associated words, with the exception of ``important''. The words ``delve'' and ``foster'' show a ten-fold increase. While the change in the prevalence of ``important'' was not statistically significant, the observed decrease would be in line with the previous results.

\begin{table}[ht!]
    \centering
    \caption{Prevalence of key stylistic terms before and at least one year after the release of ChatGPT. `Change' indicates the magnitude of change (e.g., the relative usage of the term `delve' increased 10.45-fold). $p_{MWU}$ is the p-value from a Mann-Whitney U test.}
    \begin{tabular}{lrrrr}
        \hline
        Word & Pre-ChatGPT & Post-ChatGPT & Change & $p_{MWU}$ \\ 
        \hline
        delve & 0.00000277 & 0.00002897 & 10.45$\times$ & < 0.00001 \\ 
        crucial & 0.00011832 & 0.00090101 & 7.61$\times$ & < 0.00001 \\
        potential & 0.00066499 & 0.00176767 & 2.66$\times$ & < 0.00001 \\
        these & 0.00213032 & 0.00276834 & 1.30$\times$ & < 0.00001 \\
        significant & 0.00025782 & 0.00079693 & 3.09$\times$ &< 0.00001 \\
        important & 0.00120987 & 0.00098560 & 0.81$\times$ &  0.437 \\
        leverage & 0.00001966 & 0.00007451 & 3.79$\times$ & < 0.00001 \\
        foster & 0.00005565 & 0.00060319 & 10.84$\times$ & < 0.00001 \\
        critical & 0.00012611 & 0.00051758 & 4.10$\times$ & < 0.00001 \\
        \hline
    \end{tabular}
    
    \label{tab:llm-words}
\end{table}

\subsection{Topics of Discussion}

In terms of words relating to LLMs and GAI, we, as expected, see large increases for various related expressions (table omitted for brevity). For example, the relative prevalence of the term ``Generative AI'' increased by 53.53$\times$ (from 0.00000073 to 0.00003908; MWU p-value < 0.0001). Other terms that saw statistically significant increases include ``hallucination'' (0 to 0.00000249, p = 0.0050), ``LLM'' (2.82$\times$ from 0.00001415 to 0.00003994, p < 0.0001), ``language model'' (6.21$\times$ from 0.00000301 to 0.00001868, p < 0.0001), ``Chatbot'' (3.04$\times$ from 0.00006354 to 0.00019285, p < 0.0001), ``ChatGPT'' (0 to 0.00005854, p < 0.0001) and ``GPT-4.0'' (0 to 0.00000358, p < 0.0001). At the same time, notable terms that did not see statistically significant change in prevalance include ``prompt engineering'' (p=0.3213), which had zero occurences in the Pre-ChatGPT data, and a minimal (0.00000014) prevalence in the later data, ``CoPilot'' (increased by 1.31$\times$, p=0.0661) and ``Midjourney'' (increased by 2.63$\times$, p=0.0617).

Table~\ref{tab:classic-terms} shows changes in the prevalence of a subset of terms associated with other types of AI systems. The data show increases for terms associated with natural language, potentially explained by interest and exposure towards LLMs, rather than NLP on its own. We also see significant drops for some high-level terms such as ``classification'' and ``reasoning'', but the opposite for, e.g., ``planning''. Interestingly, while ``recommender system'' saw a significant drop, ``recommendation system'' increased; we return to this observation in the discussion. 

\begin{table}[ht!]
    \centering
    \caption{Change in prevalence of terms associated with `classic' AI and AI methods.}
    \begin{tabular}{lrrrrr}
        \hline
        Term & Pre-ChatGPT & Post-ChatGPT & Change & $p_{MWU}$  \\ \hline
        Classification & 0.00004017 & 0.00000996 & 0.25$\times$ & <0.0001 \\
        Natural Language Processing & 0.00001222 & 0.00006584 & 5.39$\times$ & <0.0001 \\
        NLP & 0.00000770 & 0.00007131 & 9.27$\times$ & <0.0001 \\
        Planning & 0.00013896 & 0.00027207 & 1.96$\times$ & <0.0001 \\
        Scheduling & 0.00001516 & 0.00006301 & 4.16$\times$ & <0.0001 \\
        Recommendation system & 0.00001194 & 0.00004860 & 4.07$\times$ & <0.0001 \\
        Robotics & 0.00006497 & 0.00010346 & 1.59$\times$ & <0.0001 \\
        Recommender system & 0.00000581 & 0.00000089 & 0.15$\times$ & 0.0004 \\
        Search methodologies & 0.00000000 & 0.00000144 & -- & 0.0006 \\
        Feature selection & 0.00000043 & 0.00000156 & 3.66$\times$ & 0.0018 \\
        Reinforcement learning & 0.00006008 & 0.00008647 & 1.44$\times$ & 0.0039 \\
        Regression & 0.00000402 & 0.00000048 & 0.12$\times$ & 0.0042 \\
        Reasoning & 0.00003386 & 0.00001952 & 0.58$\times$ & 0.0277 \\
        \hline
    \end{tabular}
    \label{tab:classic-terms}
\end{table}

Table~\ref{tab:ethics-terms} shows changes in a subset of the terms relating to ethical aspects of AI. We see statistically significant, but not necessarily huge in the absolute sense, increases for terms relating to biases of various types. While (non)-discrimination stood largely static over time, it remains a common term in absolute numbers. We also note increases especially for terms such as transparency, robustness, accountability and intellectual property (but curiously not ``copyright'')---all of which are associated with commonly identified issues relating to LLMs. Terms relating to sustainability and energy also saw some increase. There is also, understandably, increasing focus on the recent EU AI Act, while the comparatively older GDPR remained stable. We also note that ``deep fake'' and ``fake news'' saw significant decreases in prevalence, falling to 0.27$\times$ and 0.39$\times$ of what they were pre-ChatGPT. At the same time, the term ``misinformation'' increased in prevalence, potentially indicating that while specific terms have fallen out of favor, the general problem is still considered relevant.

\begin{table}[ht!]
    \centering
    \caption{Change in prevalence of terms associated with AI ethics.}
    \begin{tabular}{lrrrrr}
        \hline
        Term & Pre-ChatGPT & Post-ChatGPT & Change & $p_{MWU}$  \\ 
        \hline
        Bias & 0.00199547 & 0.00331091 & 1.66$\times$ & <0.0001 \\
        Gender Bias & 0.00001040 & 0.00001769 & 1.70$\times$ & <0.0001 \\
        Racial Bias & 0.00001678 & 0.00003002 & 1.79$\times$ & <0.0001 \\
        Diversity & 0.00019248 & 0.00029208 & 1.52$\times$ & <0.0001 \\
        (Non-)Discrimination & 0.00132948 & 0.00148499 & 1.12$\times$ & <0.0001 \\
        Fair(ness) & 0.00085475 & 0.00208319 & 2.44$\times$ & <0.0001 \\
        \hline
        Accountability & 0.00031987 & 0.00097174 & 3.04$\times$ & <0.0001 \\
        Privacy & 0.00116348 & 0.00202982 & 1.74$\times$ & <0.0001 \\
        Responsibility & 0.00201205 & 0.00225469 & 1.12$\times$ & <0.0001 \\
        Robustness & 0.00000945 & 0.00003927 & 4.16$\times$ & <0.0001 \\
        Safe(ty) & 0.00104300 & 0.00161408 & 1.55$\times$ & <0.0001 \\
        Transparency & 0.00024500 & 0.00106603 & 4.35$\times$ & <0.0001 \\
        \hline
        Copyright & 0.00004325 & 0.00004530 & 1.05$\times$ & <0.0001 \\
        Governance & 0.00006099 & 0.00022339 & 3.66$\times$ & <0.0001 \\
        Intellectual Property & 0.00002170 & 0.00007465 & 3.44$\times$ & <0.0001 \\
        AI Act & 0.00001201 & 0.00003333 & 2.77$\times$ & <0.0001 \\
        GDPR & 0.00007993 & 0.00009270 & 1.16$\times$ & <0.0001 \\
        \hline
        Sustainability & 0.00009104 & 0.00021058 & 2.31$\times$ & <0.0001 \\
        Power & 0.00052693 & 0.00144541 & 2.74$\times$ & <0.0001 \\
        Energy & 0.00020843 & 0.00031427 & 1.51$\times$ & <0.0001 \\
        Electricity & 0.00004954 & 0.00001975 & 0.40$\times$ & 0.0011 \\  
        \hline
        Deep Fake & 0.00002804 & 0.00000770 & 0.27$\times$ & <0.0001 \\
        Fake News & 0.00003751 & 0.00001429 & 0.38$\times$ & 0.0032 \\
        Misinformation & 0.00002552 & 0.00006796 & 2.66$\times$ & <0.0001 \\
        Propaganda & 0.00001343 & 0.00001134 & 0.84$\times$ & 0.5841 \\

        
        %
        
        \hline
    \end{tabular}
    \label{tab:ethics-terms}
\end{table}

Finally, we trained both normal and dynamic topic models using the Gensim Python package for essays from two specific assignments with the highest amount of student freedom on what to discuss. We do not observe any meaningful changes in the broader discussed topics before and after the emergence of ChatGPT.

\section{Discussion}



\subsection{Shifts in Language and Topics}

Our results show that the essay responses have, on average, become longer with a smaller type-token ratio since the release of ChatGPT. Prior studies that have compared human- and LLM-produced texts have highlighted that LLMs tending to produce longer texts than humans~\cite{denny2023can,safi2023work} and that the type-token ratio in LLM-produced texts is typically smaller than in texts produced by humans~\cite{andre2023detecting}. 

Our results also confirm observations from prior work that has highlighted the increased use of certain LLM-specific words such as ``delve''~\cite{kobaketal2024}. An open question, however, is to what extent the use of LLMs has already influenced language use, making what we consider ``LLM-indicator'' terms equivalently common in truly human-produced language. This could happen, e.g., where students extensively use LLMs as an aid while learning English, and thus pick up the idiosyncrasies of LLMs to use in their own writing as well \cite{reviriego2024}. Studies show that non-native speakers tend to use GAI tools more often than native speakers when writing texts in English~\cite{kobaketal2024}. 

We observed significant changes in the prevalence of key terminology on AI technologies and ethics, but failed to find observable changes using topic modeling. Care should be taken when interpreting the result, especially in terms of topic modeling, as the essay prompts have stayed constant. The shift towards discussing language, LLMs, and LLM-specific concerns is also understandable, as LLMs and GAI have been the focus of public discourse. Similar shifts in technical terminology have occurred before \cite{haenlein2019,reed2014}. We are, however, somewhat concerned that the term ``AI'' will increasingly become a synonym for LLMs. 

The changes in terminology in focus could derive from at least two causes: shifting focuses of interest and public discourse applied to even a static essay prompt (e.g., when asked what they view as a key issue); or, more pessimistically, from the increased use of LLMs to write essays, and said LLMs emphasizing other topics than the average human student. Our data do not allow us to quantify whether the latter is a real phenomenon, or, if so, how significant.

Interestingly, we also saw at least two cases where terminology seems to be standardizing: ``intellectual property'' saw a 3-fold increase, while ``copyright'' remained constant, and ``recommendation system'' saw a 4-fold increase while ``recommender system'' fell to 0.15$\times$ of what it was pre-ChatGPT. Our analysis does not allow us to conclude the cause, but one possibility is that it is an artifact of increasing LLM use, which would presumably favor one of multiple near-synonyms over the other.

\subsection{Implications for Academic Integrity}

While our data cannot rule out a coincidence, nor can we conclusively point at any specific essay as being AI-generated, we view especially the essay length data as strongly suggesting that a meaningful proportion of our students have started to submit at least partially LLM-generated essay answers. This raises natural questions of whether MOOCs based on peer- or faculty-reviewed essay submissions are suitable in a context where academic (dis)honesty matters, such as when institutes want certificates of MOOC completion to hold value, unless significant breakthroughs in LLM-detection are made or MOOC-providers implement counter measures, such as relatively invasive and costly proctored online exams. 

We do note, however, that the context of our study (a free MOOC) might be more susceptible to this type of academic dishonesty than, e.g., courses taken as part of a degree. As such, things might not necessarily be quite as bleak for degree programs that are also better equipped to, for instance, include online proctored or on-premises exams.

A further question is whether the (likely) signs of LLM usage observed in our data indicate that students are actually having LLMs produce their answers wholesale, or whether they first write their response in, e.g., a non-English language and then use LLMs to translate the text. While stylistic vocabulary might be explained by the latter, the changes in essay lengths lead us to believe the former is likely more often the case.

On a broader note, our data also support the notion that the availability of GAI tools can trigger ``metacognitive laziness''~\cite{fan2024beware}, where learners offload metacognitive tasks to these tools, spending less effort on learning-related tasks. The dependence on AI tools is also linked with lower critical thinking skills, while critical thinking skills are associated with higher educational attainment~\cite{gerlich2025ai}. Similarly, confidence in GenAI is linked to critical thinking, with greater confidence in GenAI reducing it and higher self-confidence enhancing it, while shifting critical thinking toward tasks such as verification and integration \cite{lee2025}.





\subsection{Limitations}

Our work comes with some important limitations. First, we lack demographic information about the study participants. Furthermore, our data come from a MOOC on a specific topic, and our results do not necessarily extend to other topics, demographics, or forms of education. We especially caution against viewing our results as extending to all university-level teaching.

There are further limitations regarding the analysis of topics and topical terms specifically. First, it is possible that our term lists missed some key terms that would also have exhibited interesting changes. Also, due to space concerns, we were only able to report on a subset of the word lists we believed to be the most interesting. Thus, the results should not be viewed as a full and complete picture, but rather as descriptive indications of what areas might be of interest for more detailed future study. Finally, while we did not identify meaningful changes using (dynamic) topic models, this does not necessarily mean that such changes do not exist---absence of evidence is not (necessarily) evidence of absence.

\section{Conclusions}

In this work, we studied how student-submitted essays have changed following the release of ChatGPT, in terms of RQ1) style and length of writing, RQ2) non-topic related vocabulary, and RQ3) essay topics. We identified significant and persistent changes in terms of increasing essay lengths, lower text complexity, smaller variability in vocabulary and increased (relative) prevalence of terms associated with the use of LLMs. While we did observe statistically significant changes in the (relative) prevalence of individual essay topic-relevant terms relating to AI, LLMs and the ethics of AI, we did not observe meaningful changes using (dynamic) topic models.

Although we cannot explicitly confirm that students have used these tools, we find that the data strongly suggest that a meaningful proportion of MOOC participants rely on LLMs such as ChatGPT to produce essay answers. This raises important questions about the value of MOOC completions and certificates in a post-ChatGPT world. Further, our data highlight a potentially concerning trend in which students increasingly offload learning tasks to LLMs, reducing the metacognitive effort associated with learning.




%
%
%
\bibliography{ethics-of-ai-arxiv}
%

\end{document}